\documentclass[prb,preprint]{revtex4}

\pdfoutput=1

\begin{document}

\title{Comment on ``An algebra and trigonometry–based proof of Kepler’s first law'' by Akarsh Simha, arXiv:2111.08447}


\medskip 

\date{November 18, 2021} \bigskip

\author{Manfred Bucher \\}
\affiliation{\text{\textnormal{Physics Department, California State University,}} \textnormal{Fresno,}
\textnormal{Fresno, California 93740-8031} \\}

\begin{abstract}
The recent non-calculus proof of Kepler's first law succeeds because of an obscure, but valid property of the ellipse.
\end{abstract}

\maketitle

A recent article\cite{1} provides a non-calculus proof of Kepler's first law for undergraduate physics students and amateur enthusiasts. To this end two equations are derived with the same dependence on variables---$r$ and $\phi$---along with corresponding coefficients.
In one of these equations (not shown here), planet dynamics is obtained from conservation laws (energy $E$ and angular momentum $\mathbf{L}$), expressed in terms of radial distance $r =|\mathbf{r}|$ and speed $v =|\mathbf{v}|$. A $sin\;\phi$ term stems from the vector cross product $\mathbf{L} = m \; \mathbf{r \times v}$. No speed $v$ appears in the equation due to cancellation.
The other, geometric equation expresses the triangle $FPF'$, spanned by a point $P$ on the perimeter of an ellipse and the foci $F$ and $F'$, derived with the law of cosines and trigonometric identities,
\begin{equation}
    (2ar -r^2)\;sin^2\phi = a^2(1-e^2) \;,
\end{equation}
in terms of semi-major axis $a$, excentricity $e$, and angle $\phi$ between the radius vector $\overline{FP}$ and the tangent at $P$. A comparison of the equations proves that a planet's orbit is an ellipse and a comparison of coefficients yields the ellipse parameters $a$ and  $e$ in terms of $E$, $L$, masses $M$ and $m$, and gravitational constant $G$.

While the proof's result is reassuring to an undergraduate student or enthusiast by its familiarity with textbooks, it is not obvious why it works. The author mentions that it ``employ[s] an unusual equation of the ellipse." It concerns indeed an obscure---but perfectly valid---ellipse property. It can be stated as: 

\textit{The geometric mean of the projections of $F'P = 2a-r$ and $FP = r$ onto the normal to the tangent of an ellipse at $P$ equals the length of its semi-minor axis, $b$.}

Going beyond algebraic procedure, this gives a geometric interpretation of Eq. (1). To this end we rewrite it as
\begin{equation}
    (2a -r)sin\phi \;\; rsin\phi = b^2 \;.
\end{equation}
With $sin\;\phi = cos(90^{\circ} - \phi)$, the product terms on the LHS are projections of $F'P$ and $FP$ onto the normal to the tangent at $P$.  The RHS follows from $a^2(1-e^2) = a^2 - c^2 = b^2$. 

A trivial case is at hand when point $P$ is at the intersection of the ellipse's minor axis and perimeter (obtuse vertex). The tangent is then parallel to the major axis and the normal is along the minor axis. The projections of both obliques, $F'P$ and $FP$, onto the normal coincide with the semi-minor axis, $b$. Not quite that obvious is the case when $P$ is at the intersection of the major axis and perimeter (acute vertex). Opposite to the previous case, the tangent is now parallel to the minor axis and the normal is along the major axis. The projections of the obliques are $a+c$ and $a-c$, respectively. Their product is $(a+c)(a-c) = a^2-c^2=b^2$. All other cases are too complicated to recognize the theorem by visual inspection or Eq. (2) by simple algebra. That's why this property of the ellipse is rather obscure.


\end{document}